# High-density nuclear matter with nonlocal confining solitons

Charles W. Johnson* and George Fai†

*Center for Nuclear Research, Department of Physics, Kent State University, Kent, OH 44242*

## Abstract

An infinite system of nonlocal, individually confining solitons is considered as a model of high-density nuclear matter. The soliton-lattice problem is discussed in the Wigner-Seitz approximation. The cell size is varied to study the density dependence of physical quantities of interest. A transition to a system where quarks can migrate between solitons is found. We argue that this signals quark deconfinement. The model is applied to the calculation of selected in-medium properties.

*Electronic mail (internet): johnson@ksuvxd.kent.edu.

†Electronic mail (internet): fai@ksuvxd.kent.edu.

## I. INTRODUCTION

High-density strongly-interacting matter is in the focus of attention of nuclear research for several reasons. Studying large (on the order of 100 $\mathrm{fm}^3$) volumes of dense hadronic matter experimentally is one aspect of the general effort to extend the investigation of the nuclear phase diagram beyond standard nuclear matter density and zero temperature. This experimental program is carried out at a set of accelerator facilities capable of colliding heavy nuclei at increasing energies. Such collisions provide the only way to access conditions in the laboratory that were dominant in an early stage of the evolution of the Universe and are relevant today in certain celestial objects and events, like dense stars and supernovae.

It is of particularly great interest to identify and characterize in the laboratory the transition where strongly-interacting matter no longer appears as a collection of hadrons, but as deconfined quarks and gluons in an extended space-time domain. The high-density, low-temperature region of the nuclear phase diagram is especially important, as it may provide access to deconfined quarks without involving copious particle production and other effects of high excitation. Experimental results from the Brookhaven Alternating Gradient Synchrotron (AGS) indicate that this regime may be reached in the so-called full-stopping scenario achieved at the AGS.

In the present paper we discuss a static approximation for this kind of high-density nuclear matter. Since we want to address the transition to the quark-gluon phase, we start with a description of the nucleon in terms of the underlying degrees of freedom. Ideally, such calculations should be carried out in the framework of quantum chromodynamics (QCD). However, as long as the solution of QCD at nuclear length and energy scales (nonperturbative regime) remains out of reach, modeling of QCD will play an important role. Here we focus on describing strongly-interacting matter at zero temperature, as a function of density. Soliton matter has been used to model high-density hadronic matter earlier. [1–3] Bound states in a background soliton form the basic idea for the popular description of baryons with the Skyrme lagrangian. [4]

The elementary building blocks of the model proposed here are provided by the Global Color Model (GCM). [5] The GCM admits soliton solutions with an intrinsically generated, extended $\bar{q}q$ meson field, in contrast to e.g. the Color Dielectric Model, [6] which uses an external field to generate a cavity (where quarks can propagate) in the vacuum. In addition, the individual GCM solitons are confining, as there are no poles in the quark propagator outside the region where the meson field is nonzero. [7,8] The GCM observes the global symmetries of QCD, but is not locally gauge invariant. It has been used successfully to model low-energy QCD, as illustrated by the reproduction of chiral perturbation theory results, [5,9,10] meson form factors, [11] and spectra, [12] and both, the soliton (mentioned above), and the Faddeev [13] description of the nucleon. A more exhaustive summary of these successes and of current work on hadron physics based on the GCM can be found in a recent review. [14]

To address high-density nuclear matter, we consider a lattice of GCM solitons. As this model maintains only global color symmetry, our work can be considered complementary to approaches that are concerned with a complete treatment of the color degrees of freedom for the description of nuclear matter. [15] Furthermore, the kinetic energy of the nucleons (solitons) is neglected in the present work. While this can be considered a reasonable approximation at the lowest temperatures, it implies that no quantitative agreement with e.g. the saturation density of nuclear matter should be expected. For the time being, we are more interested in identifying qualitative changes in the behavior of the system with increasing density, which can justify further work on extended strongly-interacting matter with this model. We describe the the soliton lattice at the mean field level, utilizing the Wigner-Seitz approximation. [16] As a consequence, the present study is restricted to spherically symmetric mean fields. We are going to calculate excited states with higher orbital angular momenta in the spherical background field.

The density of the system can be varied by changing the size of the Wigner-Seitz cell. This allows the study of stationary energies and in-medium properties as functions of the density in the model. Even though the simplicity of our picture precludes detailed quantitative predictions, we find a very interesting qualitative feature, the occurence of a transition from a color insulator to a color conductor at a certain density. We argue that this signals the deconfinement transition in the model. We show illustrative results, such as the axial-vector coupling constant and the correlation length for pion-like correlations as a function of density to highlight possible uses of the model.

The paper is organized as follows. In Section II we review the hadronization of QCD in the context of the GCM to set the stage for the present work and to present the coupled equations defining our numerical problem. In Section III we discuss our choice of the gluon propagator and present results on a single soliton. Section IV describes how we put the GCM "on the lattice," gives details on the solution method, and presents our results on soliton matter as a function of density, including energy bands and in-medium properties. We discuss the significance of these findings. Finally, in Section V we summarize the present status of the project and outline our future plans.

## II. FROM QCD TO THE GCM SOLITON EQUATIONS

As quarks and gluons are directly unobservable, it is natural to seek a description of low-energy strong-interaction phenomena in terms of effective hadronic degrees of freedom. An example of the successes of such modeling is provided by Quantum Hadrodynamics. [17] Ideally, the effective degrees of freedom should be derived from the QCD Lagrangian. An approach to connect QCD and effective hadronic field theories can be formulated in terms of functional integral methods. The strategy is to transform the integration variables from quark and gluon fields to hadron fields. In order to make our discussion reasonably self-contained, here we review the major steps leading to the hadronic fields that play a central role in the present work.

One particular implementation of the above ideas is in the framework of the Global Color Symmetry Model (GCM), which starts with a truncation of QCD, [5] leading to the Euclidean action

$$S[\bar{q}q] = -\int d^4x\ d^4y \left[\bar{q}(x)(\gamma\cdot\partial + m)\delta(x-y)q(y) + \frac{g^2}{2} j^a_\mu(x) D_{\mu\nu}(x-y) j^a_\nu(y)\right] . \tag{1}$$

Here $j^a_\nu(x) = \bar{q}(x)\frac{\lambda_a}{2}\gamma_\nu q(x)$ is a local quark current, with Euclidean Dirac matrices $\gamma_\nu$ and Gell-Mann matrices $\lambda_a$. The two-point gluon function, $D_{\mu\nu}$, can be considered the phenomenological input point for the model. Using a Feynman-like gauge, $D_{\mu\nu} = \delta_{\mu\nu} D(x-y)$, the gluon propagator is particularly simple, and provides a single parameter function for the GCM. In the limit $D(x-y) \to \delta(x-y)$, the GCM reduces to the local Nambu–Jona-Lasinio model. [18] In (1), $m$ is a current quark mass, which will be taken to be zero in the following. Our choice of the gluon propagator is dictated by the requirement of quark confinement, and will be detailed in Section III. The GCM has the global symmetries of QCD, but lacks local gauge invariance.

The partition function in terms of the quark degrees of freedom can be written as

$$Z = N \int \mathcal{D}\bar{q}\ \mathcal{D}q\ \ \exp(S[\bar{q}q])\ , \tag{2}$$

where the functional integration $\mathcal{D}q$ implies integration over all values of the quark fields, and $N$ is a normalization constant. To exhibit nonlocal quark-antiquark structures in the action, a Fierz reordering may be performed, [19] which transforms the current-current term in (1) as

$$\frac{1}{2}\int d^4x\ d^4y\ j^a_\mu(x) D(x-y) j^a_\mu(y) = -\frac{1}{2}\int d^4x\ d^4y\ \mathcal{J}^\theta(x,y) D(x-y) \mathcal{J}^\theta(y,x)\ . \tag{3}$$

Here, $\mathcal{J}^\theta(x,y) = \bar{q}(x)\Lambda^\theta q(y)$ can be looked upon as a bilocal current with a quark-antiquark structure and quantum numbers specified by $\theta$. The quantity $\Lambda^\theta$ is a direct product of Dirac, flavor, and color matrices, and contains both, color-singlet and color-octet terms. We focus on the color-singlet sector in this work, ignoring correlations that correspond to diquark degrees of freedom.

To cast the partition function in terms of Bose fields, auxiliary nonlocal fields, $\mathcal{B}^\theta(x,y)$, are introduced, and the partition function is multiplied by unity in the form

$$1 = N' \int \mathcal{D}\mathcal{B} \exp\left[-\int d^4x d^4y \frac{\mathcal{B}^\theta(x,y)\mathcal{B}^\theta(y,x)}{2g^2 D(x-y)}\right]\ . \tag{4}$$

After the transformation $\mathcal{B}^\theta(x,y) \to \mathcal{B}^\theta(x,y) + g^2 D(x-y)\mathcal{J}^\theta(y,x)$, the action is bilinear in terms of the quark fields and the Grassman integration can be performed. This yields the action in terms of bilocal Bose fields as

$$S[\mathcal{B}] = TrLn\ G^{-1}[\mathcal{B}] - \int d^4x\ d^4y \frac{\mathcal{B}^\theta(x,y)\mathcal{B}^\theta(y,x)}{2g^2 D(x-y)}\ , \tag{5}$$

where $G^{-1}(x,y) = (\gamma\cdot\partial + m)\delta(x-y) + \Lambda^\theta \mathcal{B}^\theta(x,y)$ is the inverse quark propagator.

The replacement of the quark fields with Bose fields is, in principle, an exact functional change of variables. Observables calculated from the partition function are not affected by the variable transformation, but are now expressed in terms of the Bose degrees of freedom, provided the entire sum over $\theta$ is kept. This is impossible in practice, and the truncation scheme used can be developed into a systematic method of approximation. To retain the chiral content of the QCD action, at least two Bose fields need to be kept (see below).

The classical vacuum configuration $\mathcal{B}_0^\theta$ is identified by $\delta S/\delta \mathcal{B}^\theta = 0$. This produces a quark self-energy, $\Sigma(x-y) = \Lambda^\theta \mathcal{B}_0^\theta(x-y)$ satisfying a Schwinger-Dyson equation. In momentum space

$$\Sigma(p) = i\gamma\cdot p[A(p^2)-1] + B(p^2) = g^2 \int \frac{d^4q}{(2\pi)^4} D(p-q) \frac{\lambda^a}{2}\gamma_\mu \frac{1}{i\gamma\cdot q + m + \Sigma(q)} \frac{\lambda^a}{2}\gamma_\mu\ . \tag{6}$$

Numerical solutions for the amplitudes $A(p^2)$ and $B(p^2)$ are now available at different levels of sophistication. [20] As detailed in Section III, our choice for this explorative study is governed by simplicity within the context of the requirement of confinement. It is important to note that the amplitude $B(p^2)$ plays a dual role in the model: it also acts as the distributed vertex for coupling the quarks to the $\bar{q}q$ Goldstone modes. [7,21]

The fluctuations $\widehat{\mathcal{B}}^\theta(x,y) = \mathcal{B}^\theta(x-y) - \mathcal{B}_0^\theta$ are identified as the propagating Bose fields. If the color-singlet scalar-isoscalar and pseudoscalar-isovector fluctuations are retained, the formalism can be adopted to the requirement of chiral invariance by the variable transformation

$$\Lambda^\theta \widehat{\mathcal{B}}^\theta(x,y) = \frac{B(r)}{f_\pi} \widehat{\chi}(R) e^{i\gamma_5\cdot\phi(R)/f_\pi}\ , \tag{7}$$

where $r = x - y$ and $R = (x + y)/2$, are relative and cm-like coordinates, respectively, $f_\pi$ is the pion decay constant, and it has been assumed that the on-shell form factor $B$ can also be used off-shell. As a further simplification, the $\phi = 0$ point on the chiral circle can be fixed. In this case the radial fluctuations away from the chiral circle coincide with the scalar-isoscalar field variable prior to the transformation. In the numerical work that follows the single fluctuation field $\widehat{\chi}$ corresponding to this situation will be used, the notation serving as a reminder for possible genaralizations to restore chiral symmetry in the numerical model.

Letting $m \to 0$, the action (up to a constant) can be written as a sum of fermionic and bosonic terms:

$$S[\mu, \widehat{\chi}] = -Tr[\ \ln G^{-1}(\mu, \widehat{\chi}) - \ln G^{-1}(0, \widehat{\chi})] + \int d^4R\ [\frac{1}{2}(\partial_\mu \widehat{\chi})^2 + U(\widehat{\chi}^2)]\ . \tag{8}$$

The chemical potential ($\mu$) dependence of the fermion term in equation (8) ensures that a meson source from the valence quarks will be generated. [7] The $U(\widehat{\chi}^2)$ term is the effective meson self-interaction. [5] For $\mu = 0$ the inverse quark propagator takes the form

$$G^{-1}(x, y) = \gamma \cdot \partial_x\ A(x - y) + f_\pi^{-1} B(x - y) \widehat{\chi} \left( \frac{x + y}{2} \right)\ , \tag{9}$$

and the saddle-point configuration turns out to be $\widehat{\chi} = f_\pi$. [22]

Since $G^{-1}(x, y)$ is time-translationally invariant, stationary eigenstates of the form $u_j(x)$ can be obtained from a self-consistent Dirac equation, which in momentum space takes the form

$$[i\gamma \cdot pA(p^2) + B(p^2)]u_j(\mathbf{p}) + f_\pi^{-1} \int \frac{d^3q}{(2\pi)^{3/2}} B \left[ \left( \frac{p + q}{2} \right)^2 \right] \chi(\mathbf{p} - \mathbf{q}) u_j(\mathbf{q}) = 0\ , \tag{10}$$

where $\mathbf{p}$ and $\mathbf{q}$ are the three-momenta corresponding to the center-of-mass and relative variables, respectively. Note that $\chi = \widehat{\chi} - f_\pi$, and as $p_4 = q_4 = i\epsilon_j$, where $\epsilon_j$ is the energy eigenvalue, the meson vertex $B$ has an energy dependence. It can also be seen that a wave-function renormalization appears with the renormalization constant $Z_j$ given by [7]

$$Z_j = -\int d^3p\ d^3q\ \bar{u}_j(\mathbf{p}) \frac{\partial G^{-1}(i\epsilon; \mathbf{p}, \mathbf{q})}{\partial \epsilon_j} u_j(\mathbf{q})\ . \tag{11}$$

The meson field equation $\frac{\delta E}{\delta \chi} = 0$ may be summarized as

$$- \nabla \chi(\mathbf{z}) + \frac{\delta U}{\delta \chi(\mathbf{z})} + Q_\chi(\mathbf{z}) = 0 \ , \tag{12}$$

with the meson source provided by the valence quarks according to

$$Q_\chi(\mathbf{z}) = \sum_j \frac{1}{f_\pi Z_j} \int d^3x \ d^3y \ \bar{u}_j(\mathbf{x}) B(-\epsilon_j^2; \mathbf{x} - \mathbf{y}) \delta \left[ \frac{\mathbf{x} + \mathbf{y}}{2} - \mathbf{z} \right] u_j(\mathbf{y}) \ . \tag{13}$$

Equations (10) and (12) form a system of coupled differential equations for the quark wave functions and the meson field, which need to be solved selfconsistently, with the appropriate boundary conditions. The different boundary conditions distinguish the single-soliton case from a lattice of solitons.

## III. SINGLE SOLITON

### A. Gluon Propagator

One can take the point of view that the gluon propagator, $D_{\mu\nu}$ in Eq. (1), represents all phenomenological input to the model. With the appropriate choice of $D$, the GCM can reproduce key features of QCD, such as confinement. Phenomenologically successful early work with the GCM [5] employed a delta-function gluon propagator,

$$g^2 D(q) = 3\pi^4 \alpha^2 \delta^{(4)}(q) \ . \tag{14}$$

The momentum-space delta function turns the rainbow approximation Schwinger-Dyson equation (6) into an algebraic equation, and the Munczek-Nemirovski quark propagator [23] results:

$$A(p^2) = \begin{cases} 2 \\ \frac{1}{2}[1 + (1 + \frac{2\alpha^2}{p^2})^{\frac{1}{2}}] \end{cases} , \ B(p^2) = \begin{cases} (\alpha^2 - 4p^2)^{\frac{1}{2}} & p^2 \leq \frac{\alpha^2}{4} \\ 0 & p^2 > \frac{\alpha^2}{4} \end{cases} \ . \tag{15}$$

Its simplicity, in addition to its confining nature, make this form of the gluon propagator particularly appealing. It has a single strength parameter, $\alpha$, which (for a fixed energy) controls the spatial extension of both $A$ and $B$: the larger $\alpha$, the more localized $A$ and $B$ are in coordinate space. The lack of solutions to the equation $p^2 + M^2(p^2) = 0$, (where $M = B/A$ is the dynamic quark mass) indicates that (15) produces quark confinement; there is no on-mass-shell point, thus the propagation of a quark in the normal vacuum is prohibited. [8]

Though phenomenologically successful, the above simple quark propagator encounters difficulties when improvements are attempted regarding analyticity. [24] Partly for this reason, recent studies in the hadronic sector moved away from the point of view of providing the input at the level of the gluon propagator and, accordingly, from the simple form (14). In order to focus on nuclear matter and to keep complications from the GCM to a minimum, we use the delta-function gluon propagator (14) throughout this work. Note also the recent observation [14] that the parametrization (15) may well express the infrared behavior of the gluon two-point function in QCD. [25]

### B. Single Soliton Results

With the self-energy amplitudes $A$ and $B$ of (15), the coupled equations (10) and (12) can be solved self-consistently. For convenience, the Dirac equation is solved as a matrix equation in momentum space, using Gauss-Legendre quadrature, while the Klein-Gordon equation is solved in coordinate space using a functional version of Newton's method. [22] In order to check our numerical procedure, we have satisfactorily compared our results on the three valence-quark soliton to the results of an earlier investigation. [7,22] In addition, we carried out calculations for a range of the parameter $\alpha$.

Assuming spherical symmetry, the quark spinors can be decomposed as

$$u(\mathbf{r}) = \begin{bmatrix} g(r) \\ i\sigma \cdot \hat{k} f(r) \end{bmatrix} \mathcal{Y}_{jl}^{m_j}(\hat{r}) \ . \tag{16}$$

Here, $\mathcal{Y}_{jl}^{m_j}(\hat{r})$ is a vector spherical harmonic, $\hat{r}$ represents a unit vector in the direction of $\mathbf{r}$, and we have suppressed the quantum-number labels on $f$, $g$, and $u$. The numerical task is now reduced to the calculation of the radial functions $g(r)$ and $f(r)$.

The radial parts of the upper and lower components of the quark wave function and the meson field ($\chi$) for $\alpha = 1.04$ GeV are shown in Fig. 1. These results should be compared to the results of [22], where $\alpha = 1.04$ GeV was chosen to fit the experimental value of $f_\pi$. For our nuclear matter studies, we find it more important to have a reasonably close correspondence to the root mean square charge radius of the proton. As discussed in Ref. [8], this does not fix $\alpha$ in lack of explicitly considering the pion field, but it appears to call for larger values of the strength parameter. With $\alpha = 1.35$ GeV we get $\langle r^2 \rangle^{1/2} = 0.67$ fm for the RMS charge radius of the proton. Assuming that the pion cloud will increase this value by about 25-30 %, we get close to the experimental value. To leave room for the uncertainty of this estimate, we consider the range $1.04 \leq \alpha \leq 1.45$. In the following figures, when $\alpha$ is fixed, we display results for $\alpha = 1.35$ GeV. In Fig. 2 we show the Dirac wave function and $\chi$ for a single soliton with this value of the strength parameter.

Figures 1 and 2 represent typical results for a single soliton. It is noteworthy that the large and small components of the quark wave functions decay faster than exponential, [22] and are essentially zero by $r = 3$ fm for the values of $\alpha$ considered. The range of $\chi$ decreases and the magnitude of $\chi$ at $r = 0$ increases as $\alpha$ increases. The changes of the $\chi$ field can be attributed to the decrease in the range of the distributed quark-meson vertex $B$ in coordinate space as $\alpha$ increases. The narrower $B$, the closer $\mathbf{x}$ and $\mathbf{y}$ have to be in Eq. (13) to give a larger value of the source term in (12), which in turn influences the gradient of the $\chi$ field: a larger source term leads to a stronger gradient (assuming no change in the self-interaction). In Fig. 3 we plot the dependence of the quark eigen energy $\epsilon$ on the single input parameter $\alpha$. The increase of $\epsilon$ can be associated with the increasing absolute value of $\chi$ at $r = 0$. The spreading of $\chi$ in momentum space causes the coupling term in (10) to decrease, making the valence quarks to be less tightly bound.

## IV. SOLITON MATTER

### A. Wigner-Seitz approximation

As a means of describing nuclear matter, we consider an infinite collection of solitons. At the lowest energies the solitons are expected to arrange themselves in a crystal lattice. [26] Accordingly, the single-quark eigenenergies will develop into energy bands. For simplicity, we assume a simple cubic crystal (sc). For a periodic lattice, the Dirac wave function must be invariant to a lattice translation, so the solutions must have the Bloch form [27]

$$u_{\mathbf{m}}^{lat}(\mathbf{r}) = u_{\mathbf{m}}(\mathbf{r})\ e^{i\mathbf{m}\cdot\mathbf{r}}\ , \tag{17}$$

where $\mathbf{m}$ is the lattice momentum and $u_{\mathbf{m}}(\mathbf{r})$ is a Dirac spinor which has the periodicity of the lattice. To solve for the Bloch functions we employ the Wigner-Seitz approximation. [16] This amounts to considering a spherical cell of radius $R$ and solving for $\mathbf{m} = 0$ in (17). The full solution $u_{\mathbf{m}}^{lat}(\mathbf{r})$ is then approximated by $u_0(\mathbf{r})\ e^{i\mathbf{m}\cdot\mathbf{r}}$. Changing the density will be implemented by varying the cell radius $R$.

The Wigner-Seitz approximation places boundary conditions on the Dirac spinors. These conditions express the requirement that the upper component of the wave function must be periodic and anti-periodic for the bottom and the top of the band, respectively. We focus attention on the lowest-energy state of the band, for which the above, together with the $r = 0$ boundary conditions, implies

$$g'(r)|_{r=R} = f(r)|_{r=R} = 0\ , \tag{18}$$

where $g(r)$ and $f(r)$ represent the radial parts of the upper and lower components of the Dirac wave function (16), respectively. In addition, the meson field solution of eq. (12), which also appears in the source term of the Dirac equation (10) must now be periodic in $r$, so that

$$\chi(r + 2R) = \chi(r)\ ;\ \chi'(r)|_{r=R} = 0\ , \tag{19}$$

where $\chi(r)$ is the radial part of the meson field.

## B. GCM on the Lattice

It is convenient to solve the Dirac equation in momentum space, while the nonlinear Klein-Gordon equation is easier to handle in coordinate space. The Dirac equation (10) can be written in coordinate space as

$$0 = \int d^3y \left\{ (-\gamma_4 \epsilon_j + \vec{\gamma} \cdot \nabla) A(\mathbf{x}-\mathbf{y}) + B(\mathbf{x}-\mathbf{y}) + \frac{1}{f_\pi} B(\mathbf{x}-\mathbf{y}) \chi\left(\frac{\mathbf{x}-\mathbf{y}}{2}\right) \right\} u_j(\mathbf{y}) \ . \quad (20)$$

We seek solutions with the boundary conditions (18). These can be incorporated using a Fourier expansion of the form

$$f(\mathbf{x}) = \sum_{n_1=-\infty}^{\infty} \sum_{n_2=-\infty}^{\infty} \sum_{n_3=-\infty}^{\infty} F(\mathbf{k}_n) \ e^{i\mathbf{k}_n \cdot \mathbf{x}} \ , \quad (21)$$

where $\mathbf{k}_n = \frac{\mathbf{n}\pi}{R}$ is a wave number vector, with $\mathbf{n} = \{n_1, n_2, n_3\}$. Expanding $A$, $B$, $u$, and the meson-field source, we integrate over $y$ and use orthonormality to get an equation for the Fourier components of the $j^{th}$ quark wave function

$$\left\{ \begin{bmatrix} (-\epsilon A(k_n) + B(k_n))\delta_{nm} & -k_n A(k_n)\delta_{nm} \\ k_n A(k_n)\delta_{nm} & (\epsilon A(k_n) + B(k_n))\delta_{nm} \end{bmatrix} + \right.$$

$$\left. 4\pi \begin{bmatrix} \sum_{m=0}^{\infty} k_m^2 V_0(k_n, k_m) & 0 \\ 0 & \sum_{m=0}^{\infty} k_m^2 V_1(k_n, k_m) \end{bmatrix} \right\} \begin{bmatrix} g_m \\ f_m \end{bmatrix} = 0 \ , \quad (22)$$

where $g_n = g(k_n)$, $f_n = f(k_n)$, and

$$V_l = \int_{-1}^{1} B\left(\frac{\mathbf{k}_n + \mathbf{k}_m}{2}\right) \chi(\mathbf{k}_n - \mathbf{k}_m) P_l(\cos\theta) d(\cos\theta) \ . \quad (23)$$

To get this final form we have written the Dirac wave functions as

$$u(\mathbf{k}_n) = \begin{bmatrix} g(k_n) \\ i\sigma \cdot \hat{k}_n f(k_n) \end{bmatrix} \mathcal{Y}_{jl}^{m_j}(\hat{k}_n) \ , \quad (24)$$

and used the spherical symmetry of the Wigner-Seitz cell. If $m = 1, 2, \ldots, M$, then equation (22) is an $2M$-by-$2M$ eigenvalue problem for the energy eigenvalue $\epsilon$. The quantity $B/A$ plays the role of a dynamic mass and the scalar part of the self-energy $B$ also acts to couple the quarks to the meson field via (23). The self-energy terms have an $\epsilon$ dependence which makes this a highly non-linear problem. The $V_l$ term represents the Legendre coefficient for the meson field in the presence of the distributed coupling $B$. One needs to solve the Dirac equation (22) and the Klein-Gordon equation for the meson field (12) selfconsistently.

### C. Details of Solution

To solve for a soliton lattice, we first pick a starting meson field and search for the lowest energy eigenvalue of Eq. (22). Technically this means finding the energy $\epsilon$ which makes the determinant of equation (22) vanish. We start at $\epsilon = 0$ and work upwards in energy until the determinant changes sign. We then use the bisection method to find the root. Care must be taken so that the initial steps are sufficiently fine in $\epsilon$ not to miss the lowest root. With the root in hand, we can solve for the Fourier components of the Dirac wave functions. For this we first perform a lower upper triangular (LU) decomposition and use inverse iteration. [28] The momentum-space meson-field source term is constructed from the Dirac wave functions and we transform the source to coordinate space for use in the nonlinear Klein-Gordon equation for the meson field (12). To solve this nonlinear equation, we treat equation (12) as a functional of $\chi$ and use Newton's method. Once the Klein-Gordon equation is solved for the new meson field, we start over with the Dirac equation in this modified meson field. We iterate until convergence of the quark wave functions is achieved, which takes between three to six iterations to reasonable accuracy.

### D. Lattice Results for the Fields

When the convergence of the Dirac and Klein-Gordon equations has been reached, we have the selfconsistent ground state quark wave functions, meson field, and quark energy at our disposal. For $R = 10$ fm and $\alpha = 1.35$ GeV a quark energy $\epsilon = 530$ MeV is obtained. This reproduces the single soliton value ($\epsilon = 537$ MeV) within 2%. Fig. 4 displays the upper and lower components of the Dirac wave functions along with the meson field for $R = 5$ fm (with $\alpha = 1.35$ GeV). This still is a relatively large cell size and the results resemble the single-soliton case in shape, but the wave functions and the meson field go to zero faster than for a single soliton (compare to Fig. 2). The wave functions are pushed inwards by the boundary: each soliton is isolated around the center of the cell. The quark eigen energy for this case is $\epsilon = 526$ MeV, slightly smaller than at $R = 10$ fm.

For Fig. 5, we have decreased the size of the Wigner-Seitz cell to $R = 1.5$ fm while keeping the strength parameter $\alpha$ unchanged. Now the value of the meson field is different from zero at the edge of the cell. The upper component of the Dirac wave function, $g(r)$, can no longer decay to zero within the cell and obey the boundary condition (18). Its value is substantial at the cell boundary. The lowest quark energy begins to increase to $\epsilon = 542$ MeV.

Fig. 6 shows a cell with radius $R = 0.85$ fm and strength parameter $\alpha = 1.35$ GeV. The meson field is non-zero at the cell boundary and $g(r)$ is becoming relatively flat. As the quarks now strongly feel the presence of the neighboring cells, they become less tightly bound, and the eigen energy increases further to $\epsilon = 621$ MeV. Comparing Figures 4, 5, and 6, we conclude that there is a systematic evolution of the solutions as the cell size gets smaller (i.e. the density becomes larger) for a fixed value of the strength parameter. One important feature is that the quark distribution near the cell boundary becomes larger with increasing density: the solitons in neighboring cells begin to communicate. To clearly display this trend, we introduce the dimensionless variable $r/R$, and plot the upper component of the quark wave function and the meson field near the cell boundary as a function of this quantity in Fig. 7. This normalized variable is best suited for comparison between different values of $R$. The relative increase of the size of the meson field and of the large component of the wave function at the edge of the cell is now obvious.

### E. Energy Bands

Each soliton of the lattice contributes one level to each energy band. [27] In the Wigner-Seitz approximation we need to calculate only the energy for the case when the crystal momentum $\mathbf{m}$ in equation (17) is zero. To approximate the top of the energy band, one may solve the equations with antiperiodic boundary conditions, [29,30] or use an estimate for the band width. [26] At the level of our computational accuracy we do not expect these two methods to give significantly different results and follow the simpler band-width calculation. We approximate the Dirac-Bloch wave function for an arbitrary crystal momentum as

$$u_{\mathbf{m}}^{lat}(\mathbf{r}) = u_0(\mathbf{r})\, e^{i\mathbf{m}\cdot\mathbf{r}} \ , \tag{25}$$

and use (25) to calculate the expectation value for the square of the Dirac Hamiltonian to estimate the lattice momentum dependence of the energy levels in the band as

$$\epsilon_m = [\epsilon_{bot}^2 + \mathbf{m}^2]^{\frac{1}{2}} \ , \tag{26}$$

where $\epsilon_{bot}$ is the energy of the bottom band.

To obtain the possible values of $\mathbf{m}$ the lattice structure needs to be specified. For a simple cubic crystal of $N$ solitons and sides of length $L = 2RN$, the allowed values of the component of the lattice momentum in the direction of any of the three axes are

$$m = 0,\ \pm\frac{2\pi}{L},\ \pm\frac{4\pi}{L}\ldots,\frac{N\pi}{L} \ , \tag{27}$$

with the top of the energy band corresponding to $m = \frac{N\pi}{L} = \frac{\pi}{2R}$. Thus for the top energy band we obtain

$$\epsilon_{top} = [\epsilon_{bot}^2 + (\frac{\pi}{2R})^2]^{\frac{1}{2}} \ . \tag{28}$$

We have performed the same estimate assuming body-centered and face-centered cubic lattices, the face-centered cubic deviating the most from the estimate (28). Different assumptions in the lattice structure introduce an uncertainty of roughly 8 % in our results for the top of the band.

In Fig. 8 we show the three lowest energy bands for $\alpha = 1.35$ MeV as a function of the density. The different symbols represent the calculated energies of the bottom of each band. On one point we indicate a typical uncertainty we associate with our computation. The main source of this uncertainty is the freedom in prescribed tolerances at different stages of the calculation. The lines across the symbols represent a polynomial fit to guide the eye and to facilitate the calculation of the top of the bands with the approximation (28).

The lowest band ($l = 0$, $l' = 1$ and $j = \frac{1}{2}$) is labeled $1s_{1/2}$. The next lowest band has nonzero orbital angular momentum in the large component of the Dirac wave function ($l = 1$, $l' = 2$, and $j = \frac{3}{2}$) and is labeled $1p_{3/2}$. The next band is again an $s$-state, corresponding to a radial excitation. For very low density ($R \longrightarrow \infty$) the energy bands shrink to single levels and in the limit reproduce the energies of a single soliton (discussed earlier). As the density increases the ground-state band develops a minium. The low-density attraction between the solitons is a consequence of the boundary conditions on the quark wave functions (18). In particular, the upper component of the quark wave function is forced to have less curvature than in the case of a single soliton, leading to a lower value of the quark kinetic energy. At higher densities, where the solitons and the quark wave functions begin to overlap, the resulting repulsion overcomes the attraction and the ground-state quark energy starts to increase. In a more complete calculation one would like to attempt to fit the energy minimum to the saturation density of nuclear matter. This, however, requires to go beyond the present mean field treatment which lacks the details of the nucleon-nucleon interaction and nucleon kinetic energies.

As the density increases, the top of the ground-state band approaches the bottom of the next unfilled band, and at $\approx 2.6\ \rho_0$ the highest energy state of the occupied ground state band intersects the bottom of the empty $1p_{3/2}$ band. At this point it becomes energetically favorable for the quark in the highest-energy state to move into the empty "conduction" band. The system goes through a transition very similar to the insulator-conductor transition in metals, and color conductivity sets in. Since in the new phase quarks are free to migrate from soliton to soliton, we identify this transition with quark deconfinement. [1]

Note in this context that we use uniformly filled bands in our calculation. Partial filling of the lowest energy band will increase the critical density in the model.

### F. In-medium properties

The surrounding dense nuclear matter may significantly change single-particle properties like masses and widths, as well as coupling constants, cross sections, and other features relevant for transport modeling. For example, dilepton production experiments at CERN [31] seem to indicate a shifting $\rho$-meson mass in medium. [32] Here we calculate the axial vector coupling constant in the model to illustrate how the density dependence of physical properties can be obtained. As another example of an in-medium property, we present the calculation of a correlation length.

The axial vector coupling constant can be calculated in momentum space. Since there are no explicit pion fields in the model, only the valence quarks contribute. In a simple approximation [22]

$$g_A = \frac{5}{3}\frac{1}{Z}\int dp\; p^2 A(-\epsilon^2 + p^2)\;\left\{ g^2(p) - \frac{1}{3} f^2(p) \right\} \quad , \tag{29}$$

where the $\frac{5}{3}$ factor is due to the summation over spin and flavor degrees of freedom. In Fig. 9 we plot $g_A$ as a function of density for several values of the strength parameter $\alpha$. [The line serves to guide the eye.] For large cell size (low density) the single soliton result [22] is approached. In general, the value of $g_A$ decreases with increasing density. This can be understood in terms of $g(p)$ becoming more and more localized in momentum space as $g(r)$ spreads out in coordinate space with decreasing cell size. The decrease continues until the critical density, which is the highest density up to which we trust our calculations based on a single cell. At the transition we physically expect a sudden increase in the value of $g_A$ to higher than its free-space value, as a consequence of the fact that the deconfined quarks sample a larger volume of phase space. The trend at higher densities that $g_A$ is smaller for smaller $\alpha$ is consistent with the narrowing of $A$ in momentum space as the strength parameter decreases.

As another illustration of the utility of the model, we construct the pion-like bilocal fluctuation field $\bar{q}\gamma_\mu\gamma_5 q$, and calculate the current-current correlation function [33]

$$\Gamma(\mathbf{r}_1, \mathbf{r}_2) = \frac{\langle\ \bar{u}(\mathbf{r}_1)\gamma_\mu\gamma_5 u(\mathbf{r}_2)\ \bar{u}(\mathbf{r}_2)\gamma_\mu\gamma_5 u(\mathbf{r}_1)\ \rangle}{\langle\ \bar{u}(\mathbf{r}_1)\gamma_\mu\gamma_5 u(\mathbf{r}_1)\ \rangle \cdot \langle\ \bar{u}(\mathbf{r}_2)\gamma_\mu\gamma_5 u(\mathbf{r}_2)\ \rangle} - 1 \quad . \tag{30}$$

To reduce the number of variables we set $\mathbf{r}_1 = \mathbf{r}$ and $\mathbf{r}_2 = 0$. Averaging over angles and using the properties of the Dirac spinors makes it possible to write the correlation function as a function of one variable, the relative distance $r$. The result is

$$\Gamma(r) = \frac{\frac{1}{3} f^2(r)}{g^2(r) - \frac{1}{3} f^2(r)} \quad , \tag{31}$$

To extract a correlation length $R_{cor}$, (31) is transformed to momentum space and we follow the methods usually applied in Hanbury-Brown and Twiss (HBT) types of analyses for bosons: the width of the momentum-space peak is inversely proportional to $R_{cor}$. [34]

In Fig. 10 we show the correlation length $R_{cor}$ as a function of density for three values of $\alpha$. [The line serves to guide the eye.] With increasing density the cells become smaller, and $R_{cor}$ decreases. At the transition density a sudden increase of the correlation length is expected, as the quarks become free to migrate from cell to cell. For small density (large cell size) there is a large spread in $R_{cor}$ as a function of $\alpha$. This can be connected to the increasing spatial localization of the solitons as $\alpha$ increases. For smaller values of $\alpha$ the solitons spread out more, so that the lower component $f(r)$ peaks at a larger distance away from the origin. For small cell size (large density) the wave function has no room in the cell to spread out, and $R_{cor}$ becomes more independent of $\alpha$.

The above examples serve to illustrate how the model can be used to discuss the density dependence of in-medium properties. We plan to calculate other hadronic observables, like in-medium masses and further correlation functions in the future.

## V. SUMMARY

We presented a generalization of the Global Color Model (GCM) to the many-soliton situation. In particular, the density dependence of the properties of an infinite system of nonlocal, confining solitons was studied in the Wigner-Seitz approximation. We found that, at a critical density, an infinite system of solitons exhibits a transition from a geometry with one soliton localized at the center of each cell to a more uniform situation where quarks can migrate across cell boundaries. We argued that this transition signals quark deconfinement in the model. We have also calculated the density dependence of the axial vector coupling constant and of a correlation length as examples of in-medium properties.

It should be kept in mind that the least elaborate one-parameter version of the GCM was used throughout this work. More realistic parametrizations of the quark self-energy functions should improve the accuracy of the description. In particular, the pion decay constant and the root mean square charge radius of the proton could be fitted simultaneously with a couple of parameters instead of just one. Such improvements in the hadronic sector can be incorporated in the model if a closer correspondence to experimental data is desired. Developments along these lines require explicit pion degrees of freedom in the model for inclusion into the calculation of the root mean square proton radius. The inclusion of explicit pions will also lead to the restauration of chiral symmetry in the model, as discussed following Eq. (7). As this would open the way for chiral calculations, it promises to be an interesting line of future development.

It would also be of interest to perform direct comparisons to QCD calculations. This can shed more light on the nature of the transition we found in the model. We consider the existence of this transition in the GCM-based soliton lattice to be our most important finding so far. We believe that this feature is sufficiently robust to survive in more refined versions of the model.

## ACKNOWLEDGMENTS

Helpful discussions with M. Frank, who participated in an earlier phase of this work, are gratefully acknowledged. This work was supported in part by the Department of Energy under Grant No. DOE/DE-FG02-86ER-40251.

## FIGURE CAPTIONS

**Fig. 1:** Upper and lower components of the quark wave function for a single soliton with the corresponding meson field plotted for $\alpha = 1.04$ GeV. The quark eigen energy is $\epsilon = 356$ MeV.

**Fig. 2:** Upper and lower components of the quark wave function for a single soliton with the corresponding meson field plotted for $\alpha = 1.35$ GeV. The quark eigen energy is $\epsilon = 537$ MeV.

**Fig. 3:** The dependence of the quark eigen energy $\epsilon$ on the strength parameter $\alpha$ for a single soliton.

**Fig. 4:** Upper and lower components of the quark wave function for the soliton lattice with the corresponding meson field for $R = 5.0$ fm, $\alpha = 1.35$ GeV. The quark eigen energy is $\epsilon = 526$ MeV.

**Fig. 5:** Upper and lower components of the quark wave function for the soliton lattice with the corresponding meson field for $R = 1.5$ fm, $\alpha = 1.35$ GeV. The quark eigen energy is $\epsilon = 542$ MeV.

**Fig. 6:** Upper and lower components of the quark wave function for the soliton lattice with the corresponding meson field plotted for $R = 0.85$ fm, $\alpha = 1.35$ GeV. The quark eigen energy is $\epsilon = 621$ MeV.

**Fig. 7:** Upper components of the quark wave functions and the corresponding meson field for the lattice case plotted against the dimensionless variable $r/R$ for $R = 1.2$, 1.3, and 1.5 fm, and $\alpha = 1.35$ GeV.

**Fig. 8:** The bottom and top energies for the lowest energy bands of the soliton lattice as a function of the density in terms of standard nuclear matter density ($\rho_0 = 0.17$ fm$^{-3}$), for $\alpha = 1.35$ GeV. The symbols represent the calculated energies of the bottom of the three lowest energy bands. For the top of the energy band we use (28) as an approximation. An illustrative error bar is included. The curves represent a polynomial fit to guide the eye.

**Fig. 9:** The axial vector coupling constant as a function of the density for $\alpha = 1.25$ GeV, 1.35 GeV, and 1.45 GeV.

**Fig. 10:** The correlation length for pion-like currents as a function of the density for $\alpha =$ 1.25 GeV, 1.35 GeV, and 1.45 GeV.

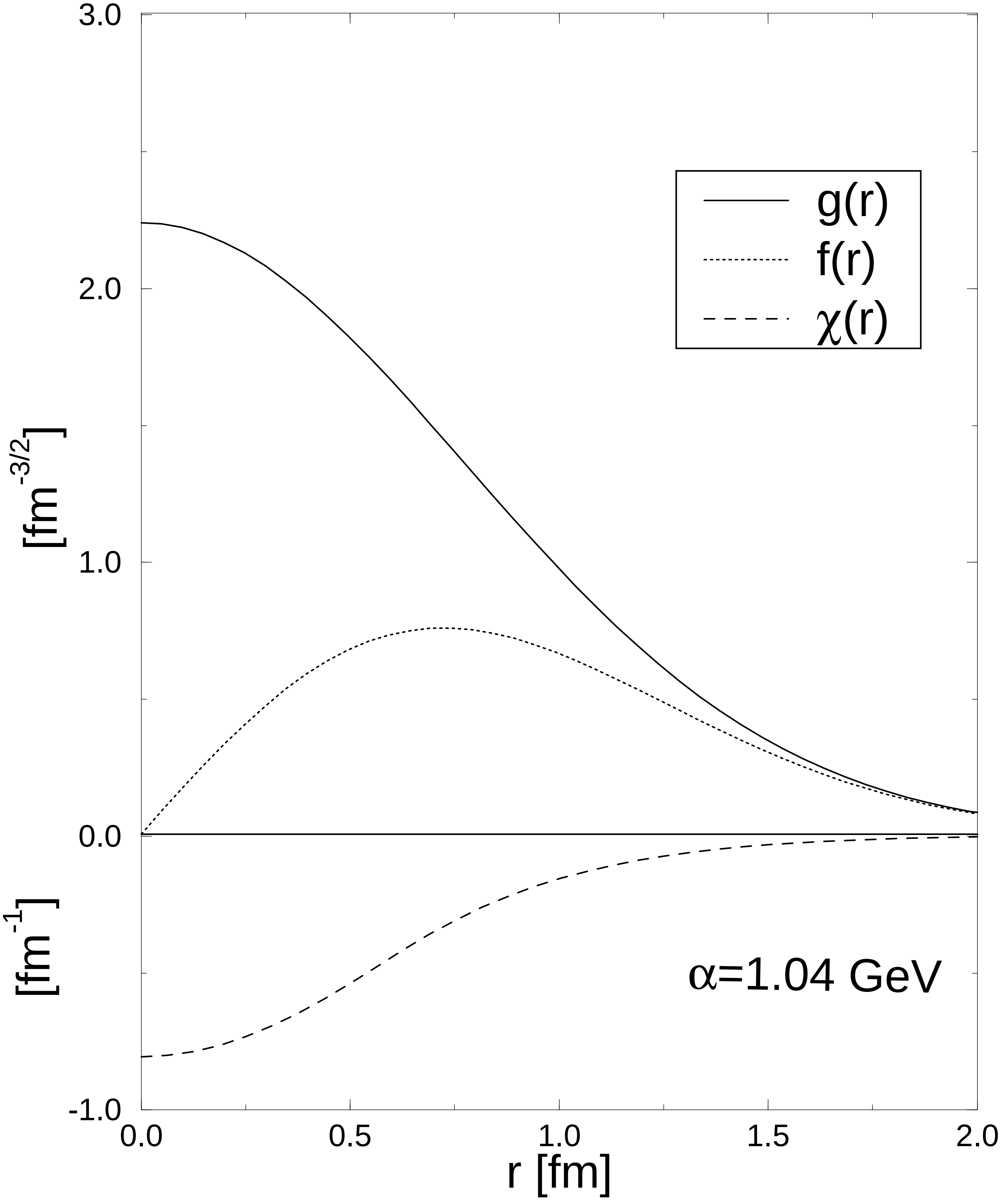

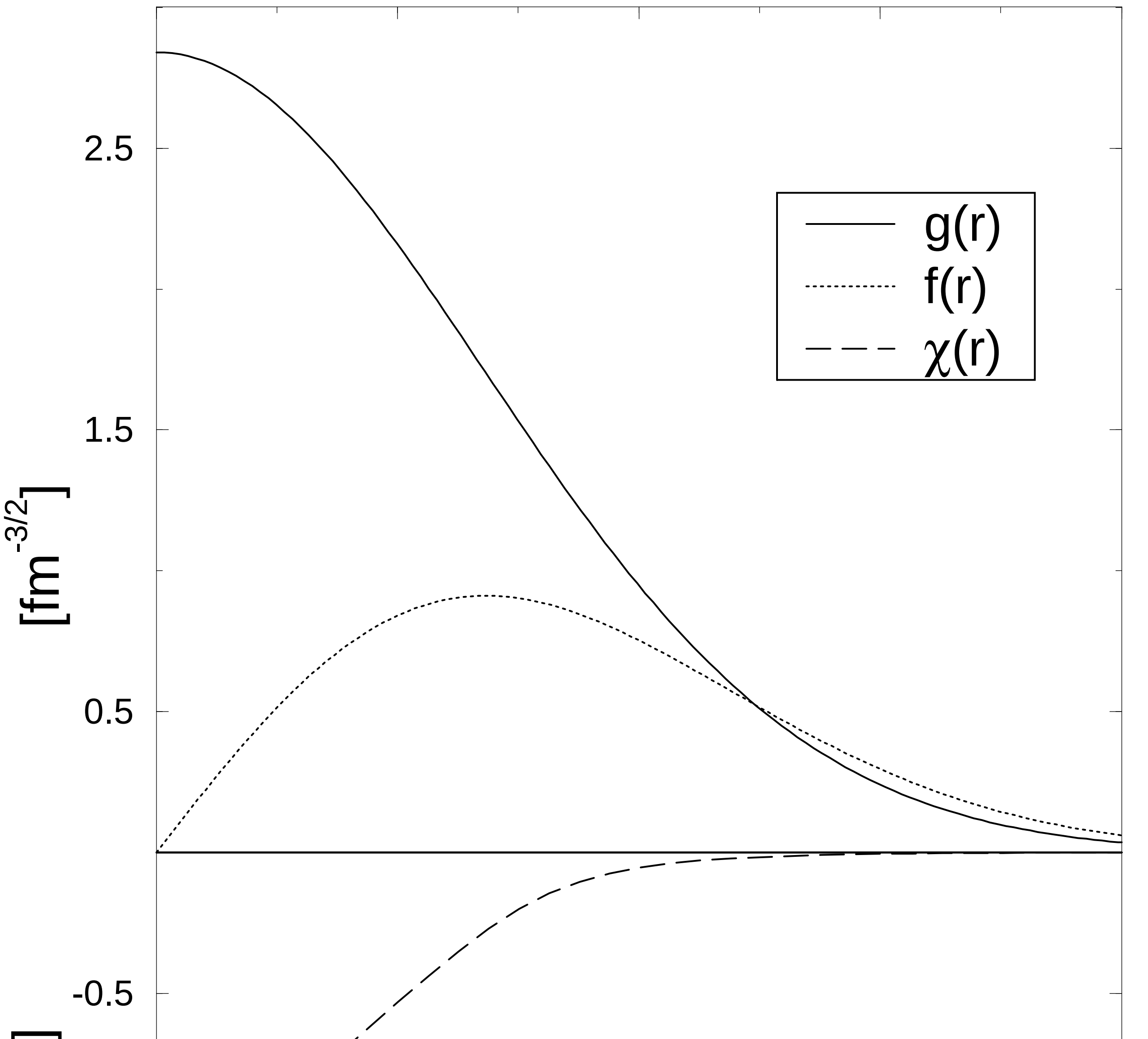

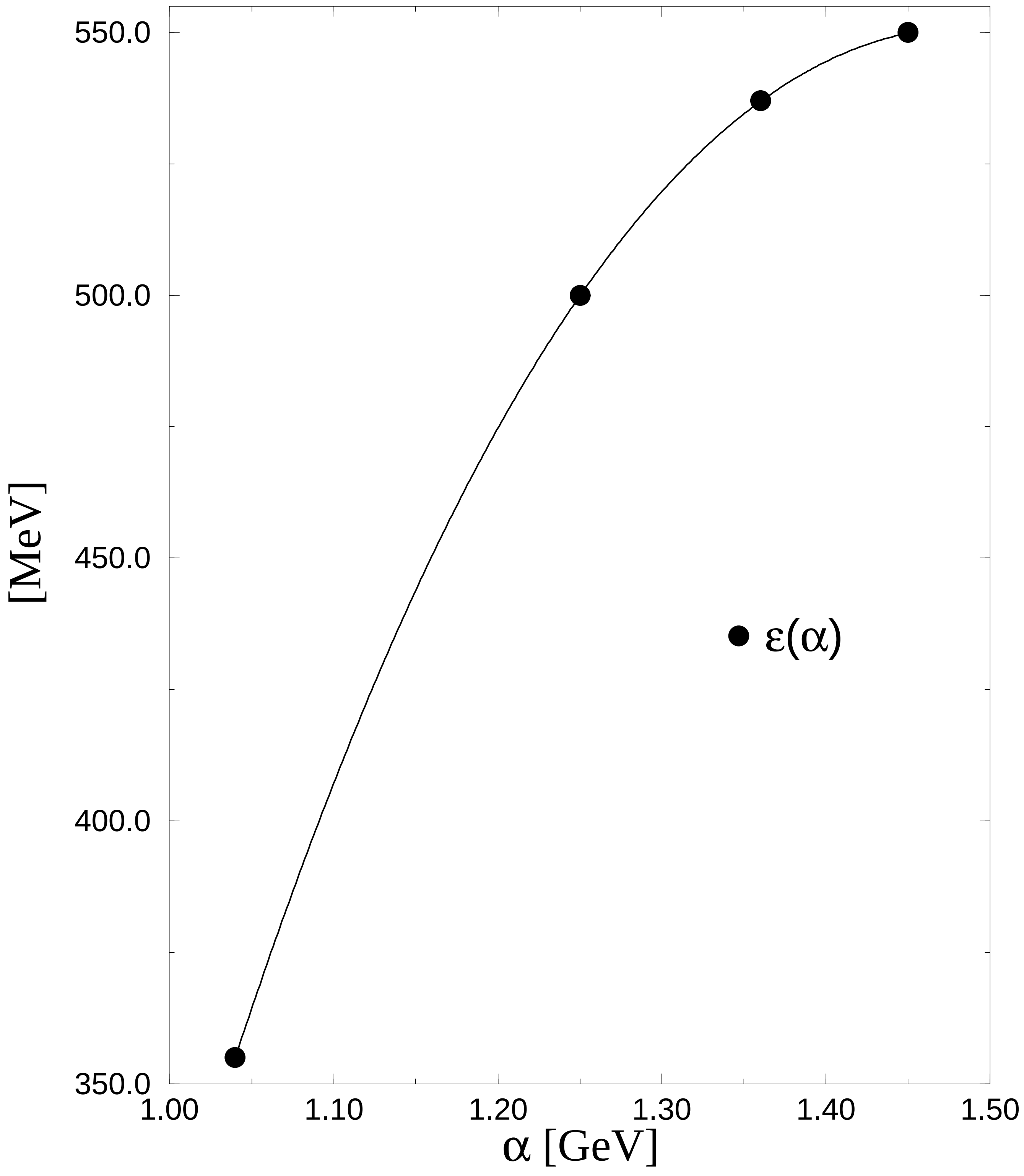

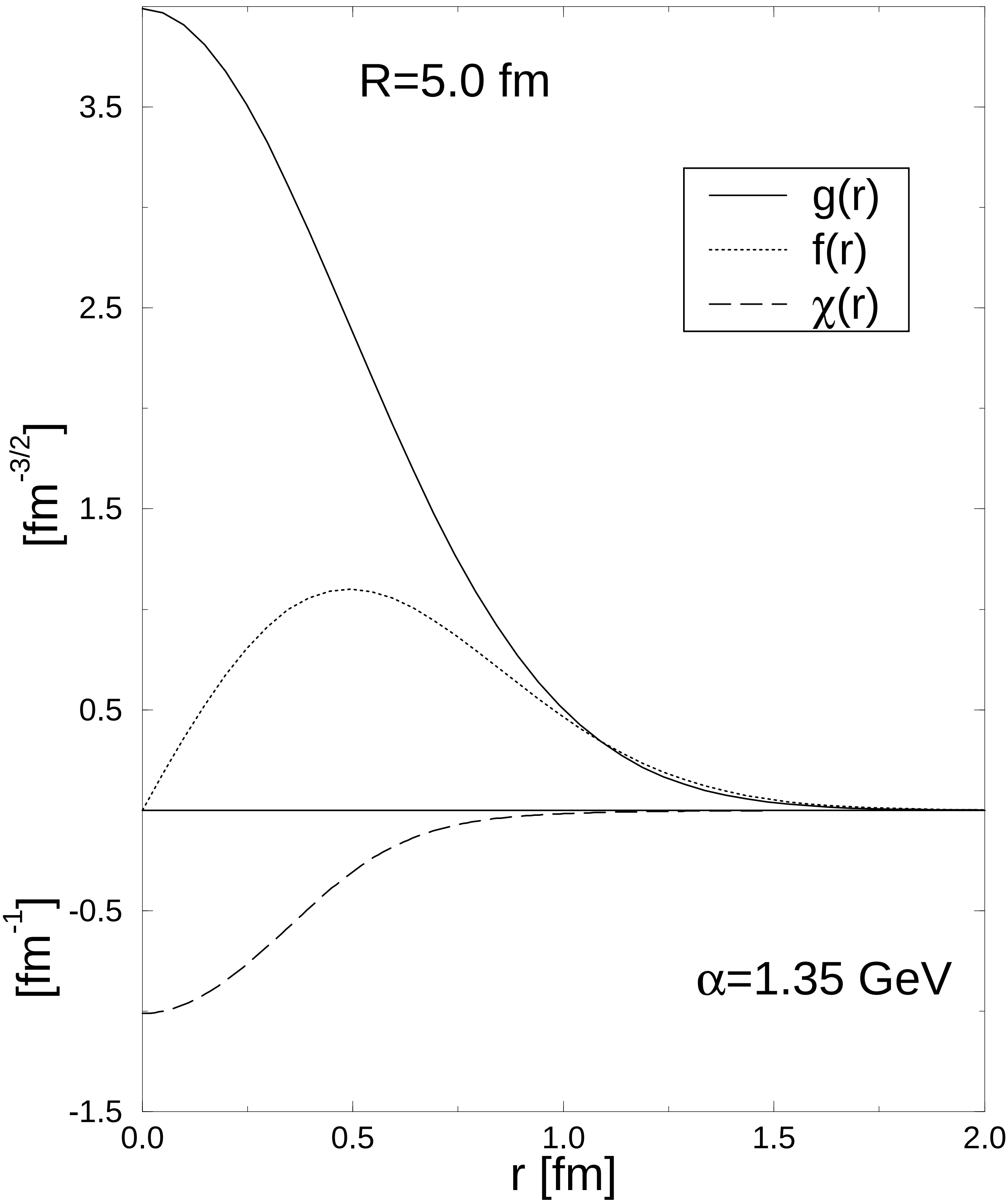

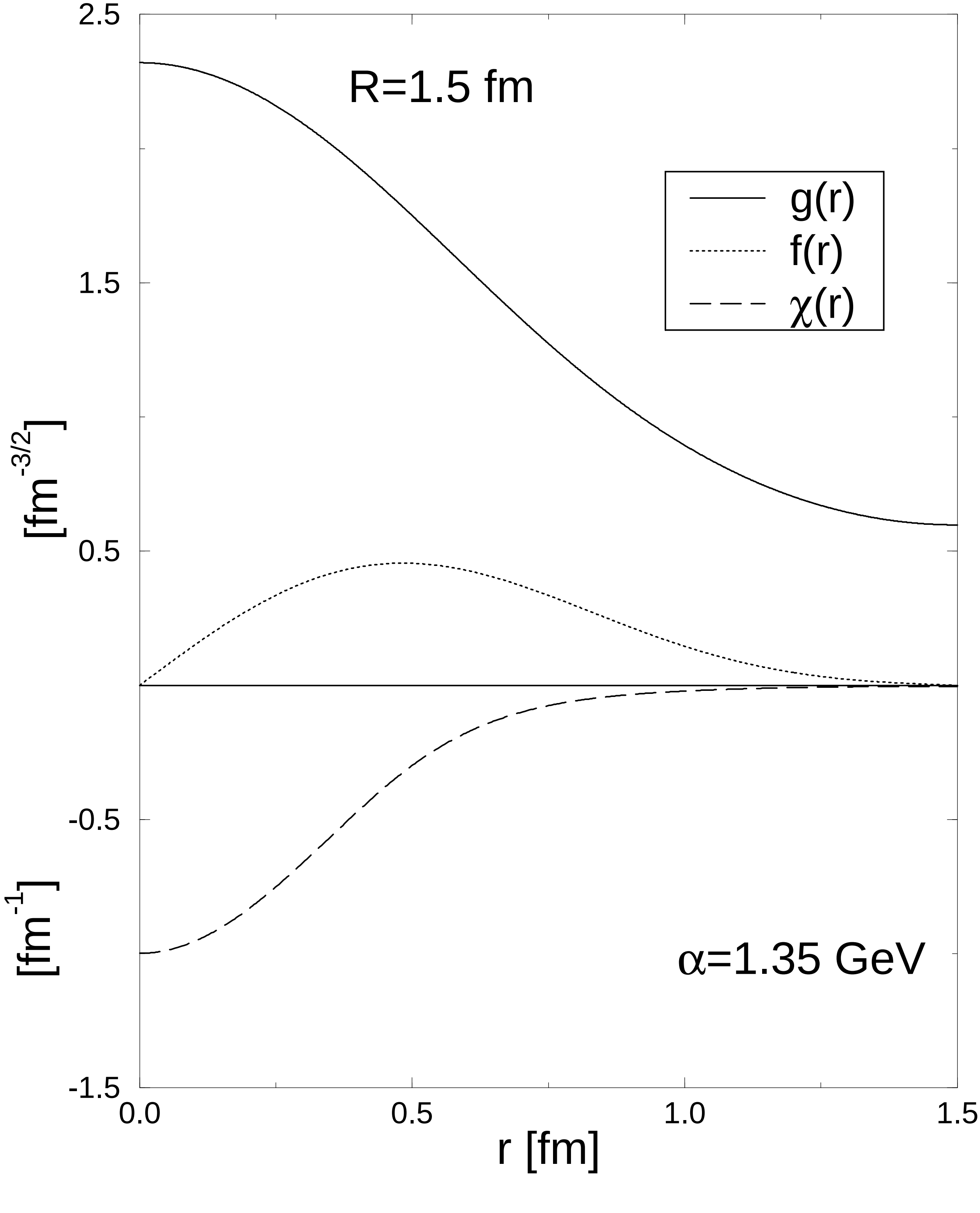

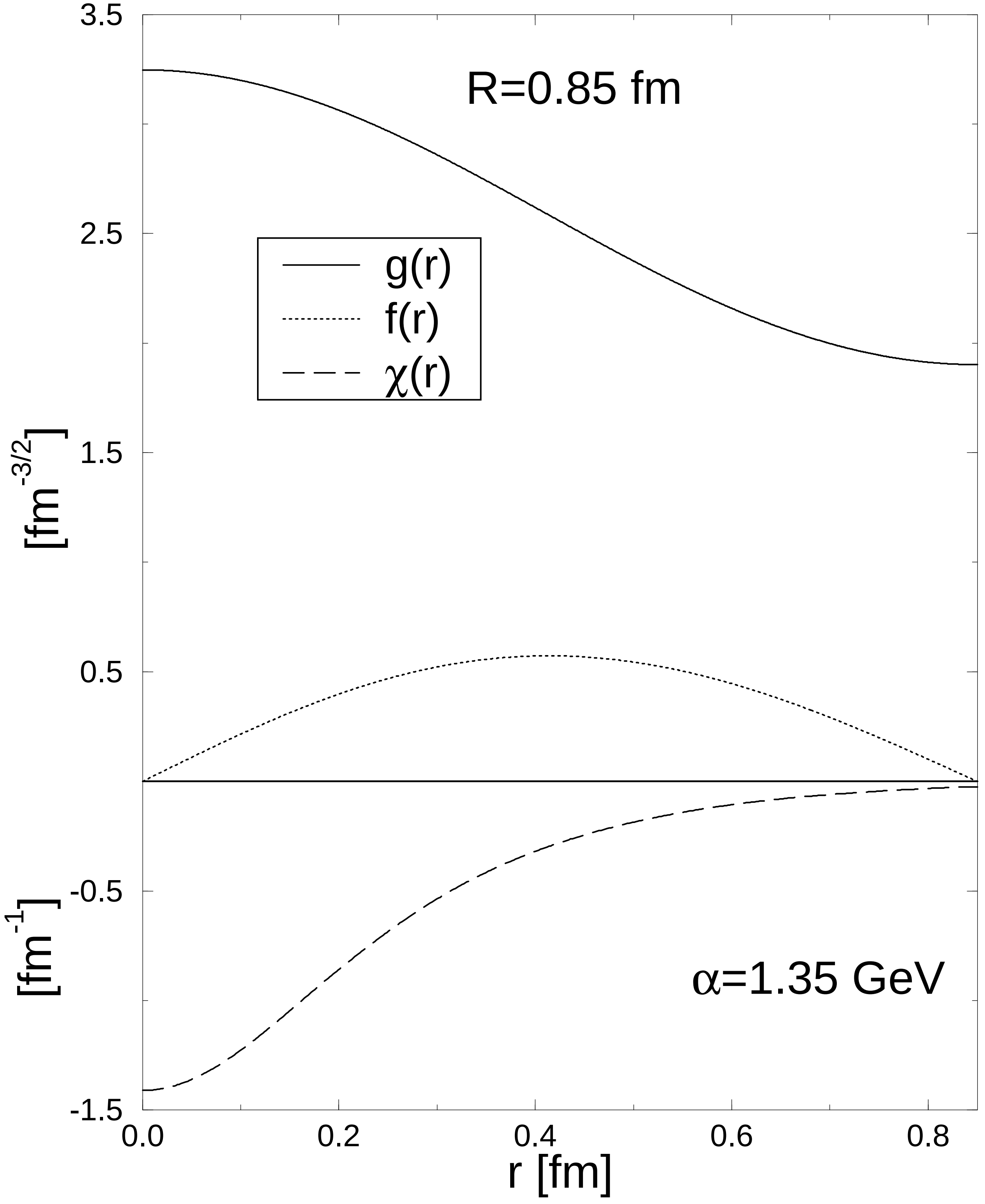

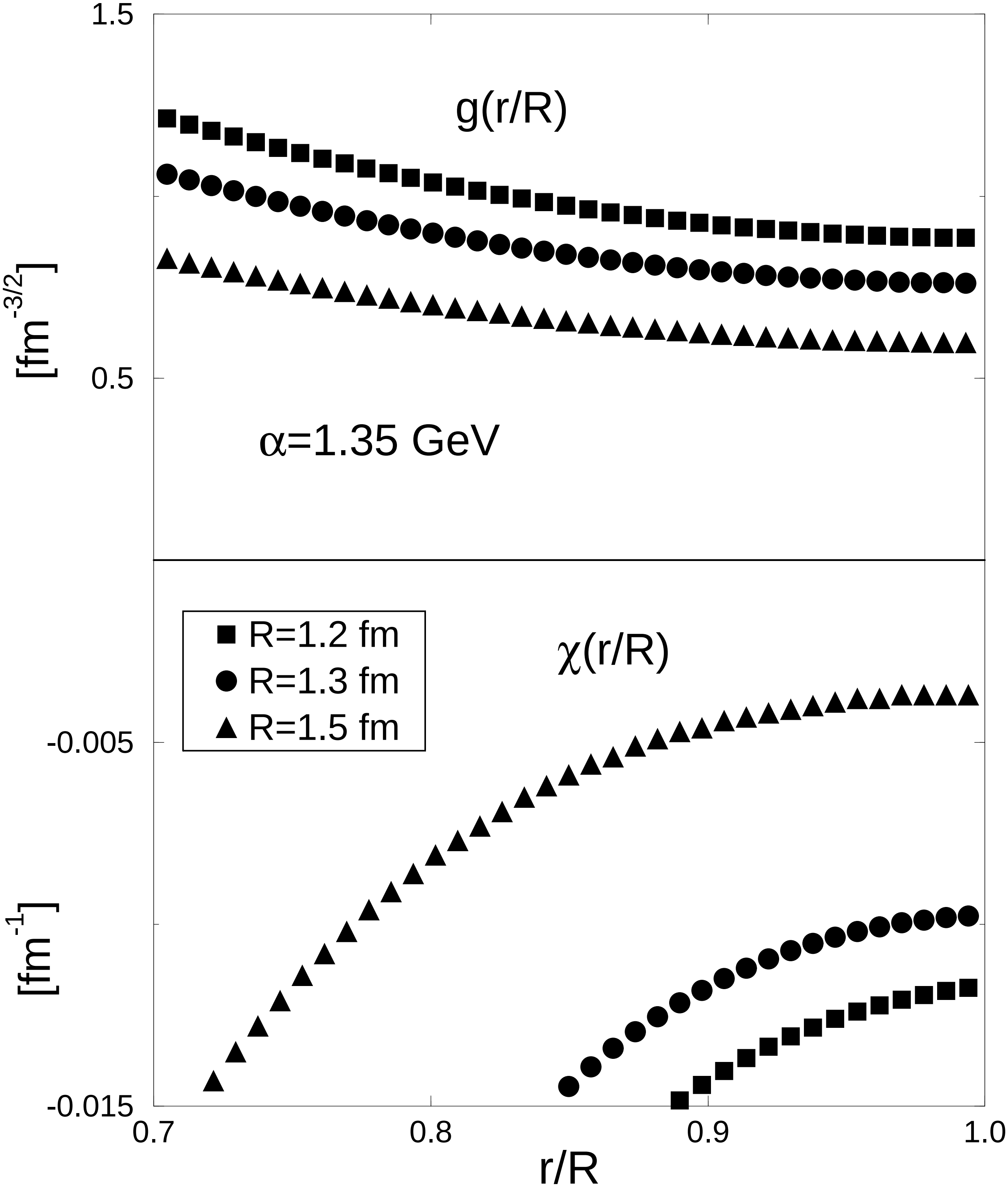

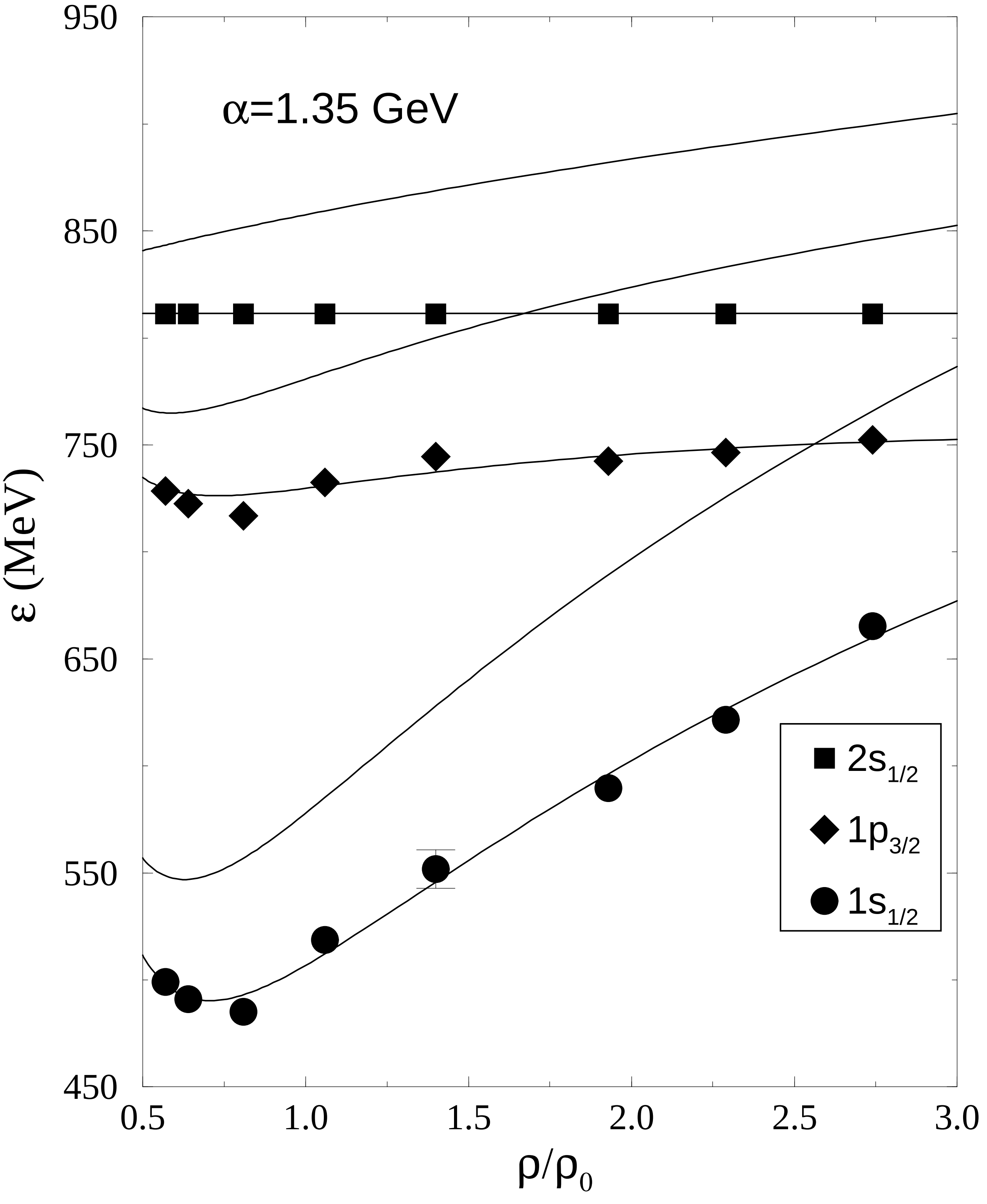

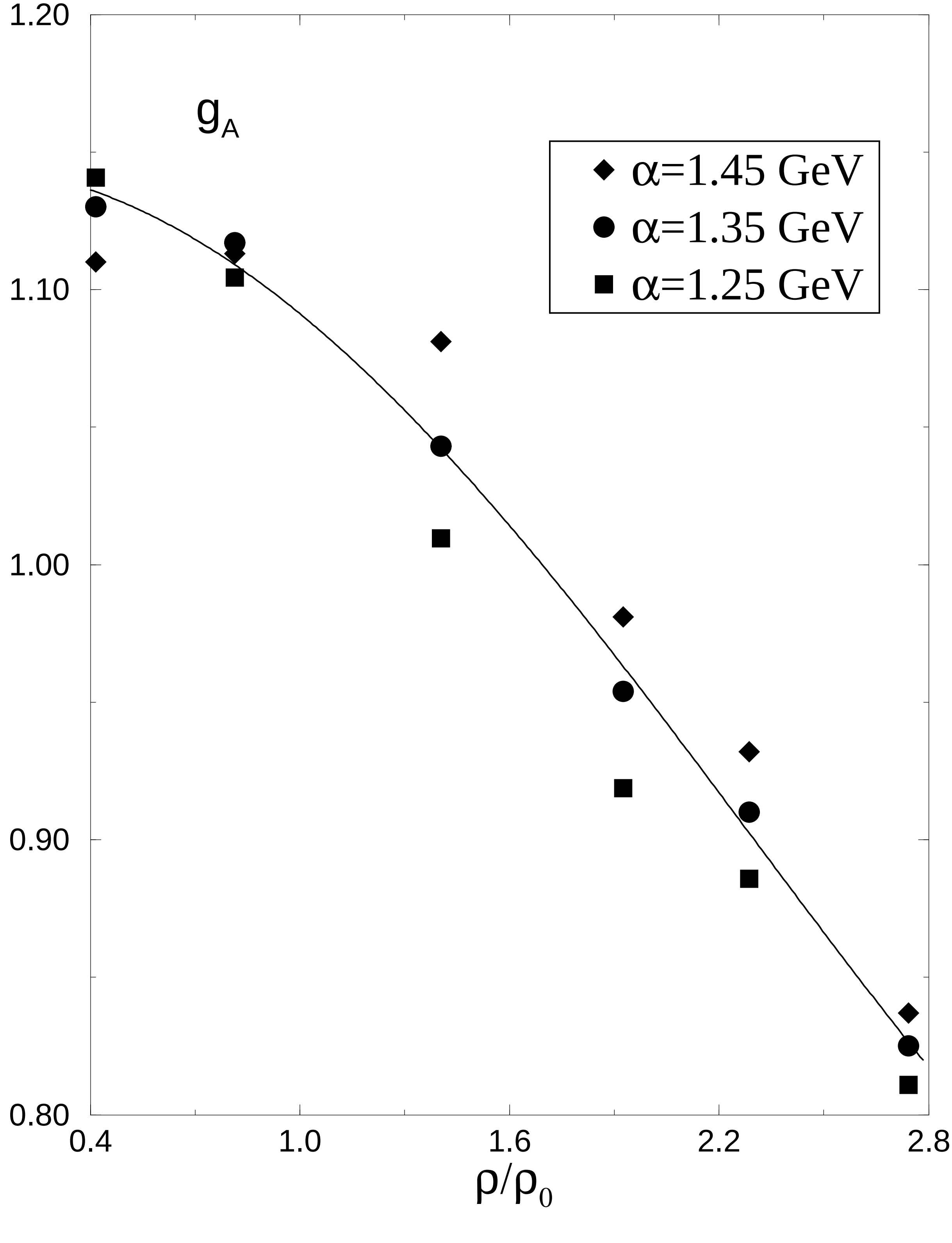

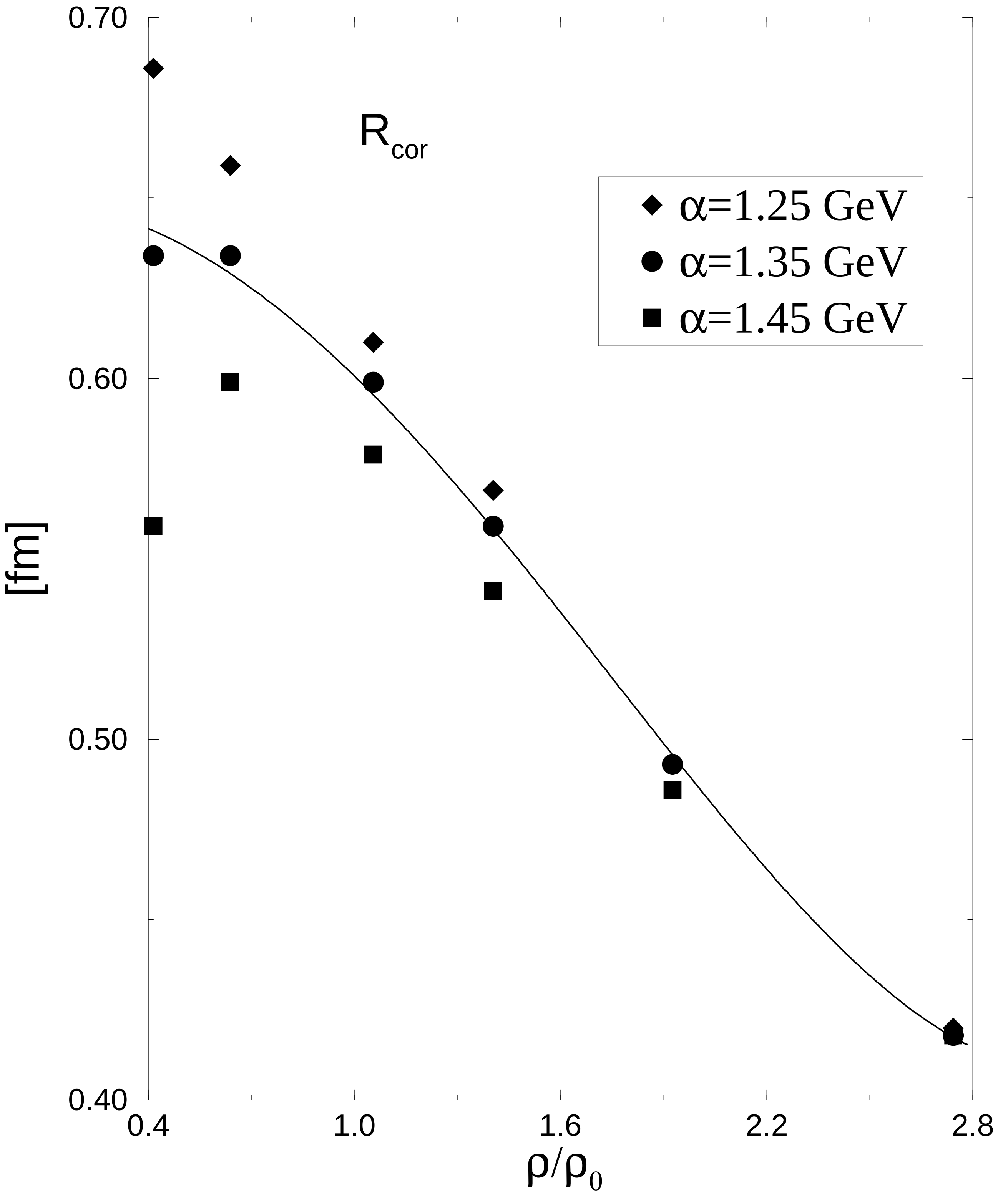